\documentclass[fleqn,usenatbib]{mnras}

\usepackage{newtxtext,newtxmath}

\usepackage[T1]{fontenc}

\DeclareRobustCommand{\VAN}[3]{#2}
\let\VANthebibliography\thebibliography
\def\thebibliography{\DeclareRobustCommand{\VAN}[3]{##3}\VANthebibliography}

\usepackage{graphicx}	
\usepackage{amsmath}	
\usepackage{threeparttable}

\newcommand{\fer}{{\it Fermi}}
\newcommand{\mx}{{\it MAXI}}
\newcommand{\sw}{{\it Swift}}

\newcommand{\egs}{\rm\,erg\,s^{-1}}
\newcommand{\hzs}{{\rm\,Hz\,s^{-1}}}

\newcommand{\cx}{Cen X-3}
\newcommand{\vx}{Vela X-1}

\title[Torque-dependent orbital modulation of Cen X-3]{Torque-dependent orbital modulation of X-ray pulsar Cen X-3}

\author[Z. Liao et al.]{
Zhenxuan Liao,$^{1}$\thanks{E-mail: liaozhenxuan@fjsmu.edu.cn}
Jiren Liu$^{2}$\thanks{E-mail: jrliu@swjtu.edu.cn}\\
$^{1}$School of Information Engineering, Sanming University, Jingdong Road 25, Sanming 365004, Fujian Province, People's Republic of China\\
$^{2}$ School of Physical Science and Technology,
Southwest Jiaotong University, Chengdu, Sichuan, People's Republic of China
}

\date{Accepted XXX. Received YYY; in original form ZZZ}

\pubyear{2023}

\begin{document}
\label{firstpage}
\pagerange{\pageref{firstpage}--\pageref{lastpage}}
\maketitle

\begin{abstract}
\cx~shows alternate spin-up/spin-down episodes lasting for tens of days.
We study the orbital profiles and spectra of \cx~during these spin-up/spin-down intervals, 
using long-term data monitored by \fer/GBM, \sw/BAT and \mx/GSC. 
In spin-up intervals, its orbital profile in 2--10\,keV is symmetrically peaked
around orbital phase 0.42,
while in spin-down intervals of similar fluxes and similar magnitudes of spin change rate,
its profile reaches a peak around orbital phase 0.22 and then declines gradually. 
Such a distinct orbital difference between spin-up and spin-down states of similar flux is hard to explain in the standard disk model and indicates that its torque reversals are related to processes on the orbital scale.
The durations of continuous spin-up/spin-down trend (tens of days) also point to a superorbital variation. 
One possible scenario is the irradiation-driven warping disk instability, which may produce a flipped inner disk for tens of days. 

\end{abstract}




\section{Introduction}

In X-ray pulsars, a magnetized neutron star accretes material from its companion and produces pulsating X-ray radiation as it rotates with a misaligned magnetic field \citep[for a recent review, see][]{Kret2019}. 
The angular momentum carried by accreting matter leads to spin frequency changes observable on a timescale from days to years. 
Many X-ray pulsars show alternate spin-up/spin-down reversals, which are not yet fully understood \citep{Bild1997,Mala2020}. 
Especially, the anti-correlation between torque and luminosity observed in the spin-down state in some sources \citep{Chak1997, Liao2022o} cannot be explained by a standard accretion disk model, 
and a variable prograde/retrograde disk scenario has been proposed \citep{Maki88,Nelson1997}.

Recently, we found that the orbital profile is different between spin-up and spin-down (torque) states in two classical X-ray pulsars, 
OAO 1657-415 and Vela X-1 \citep{Liao2022o,Liao2022v}. 
In particular, the absorption column density of Vela X-1 in the spin-up state is higher than that of the spin-down state. 
Such a torque-dependent orbital property indicates that the spin-up/spin-down state is related with the accretion flow on the orbital scale, well beyond the magnetosphere scale. 
They are consistent with a variable prograde/retrograde flow to the neutron star. 
These results motivate us to study the torque reversal of Cen X-3, which shows alternate spin-up/spin-down episodes on tens of days \citep{Finger1994,Bild1997}. 

Discovered in the early years of X-ray astronomy \citep{Chod1967}, 
\cx~is the first identified accreting X-ray pulsar, 
with a spin period of $\sim4.8\,\rm s$ and an eclipsing orbital period $\sim$2.087\,days \citep{Giacconi1971,Schreier1972}. 
\cx~has a neutron star of $1.34_{-0.14}^{+0.16}\rm\,M_{\sun}$ \citep[e.g.][]{vdm2007}, 
orbiting an O6-7III supergiant V779 Cen \citep[$\sim20.5\pm0.7\rm\,M_{\sun}$,][]{Ash1999}
with a nearly circular orbit \citep[$e<0.0003$,][]{Klawin2023}.
\cx~has long been regarded as a disk-fed SgXB, as evidenced by its variable optical light curve \citep{Tjemkes1986}.
However, no correlation was found between its spin frequency derivative and luminosity \citep{Tsunemi1996}.
Its fluxes show strong variation with an aperiodic timescale of around $\sim140$ days \citep{Priedhorsky1983}.
Its orbital modulation was found to be intensity-dependent and was suggested to be due to the varying obscuration of a precessing disk \citep{Raichur2008l}.
The torque reversals of \cx~have been suggested to be related to the irradiation-driven warping instability of the accretion disk, the inner part of which could be titled more than 90 degrees \citep{vanKerkwijk1998}.

\section{Observations}
\begin{figure*}
    \centering
    \includegraphics[width=1.5\columnwidth]{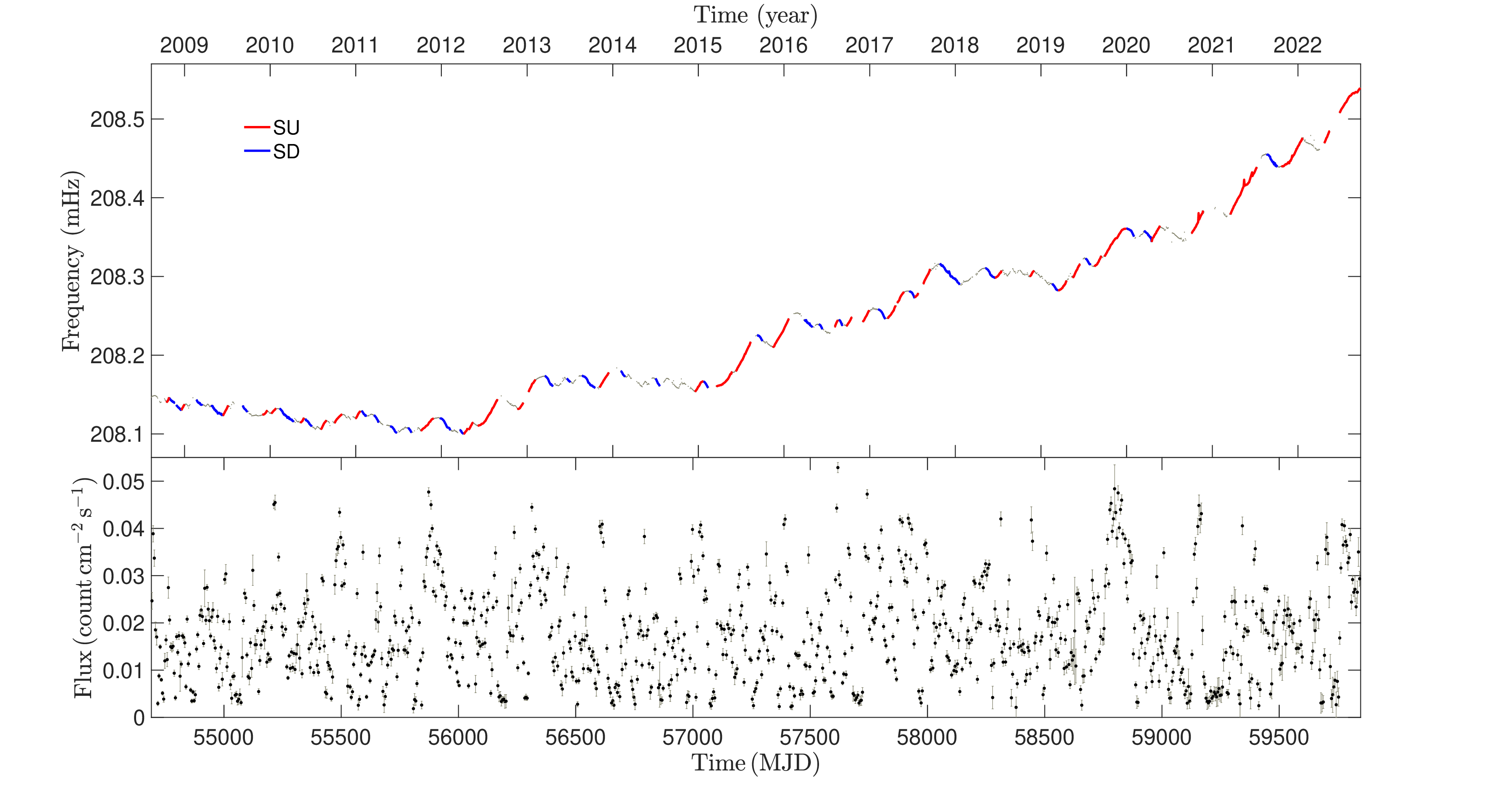}
    \caption{Upper panel: decade-long spin frequency history of \cx~monitored 
	 by \fer/GBM, with the marked spin-up (SU, red) and spin-down (SD, blue) intervals for further analysis. The errors of spin frequency measurements are of the order of $10^{-8}$--$10^{-7}\,\rm Hz$, too small to be displayed. Lower panel: corresponding flux within the 15--50\,keV band extracted from {\it Swift}/BAT, binned in 5\,days.}
    \label{fig:sh}
\end{figure*}
\begin{figure}
    \flushleft
    \includegraphics[width=1.0\columnwidth]{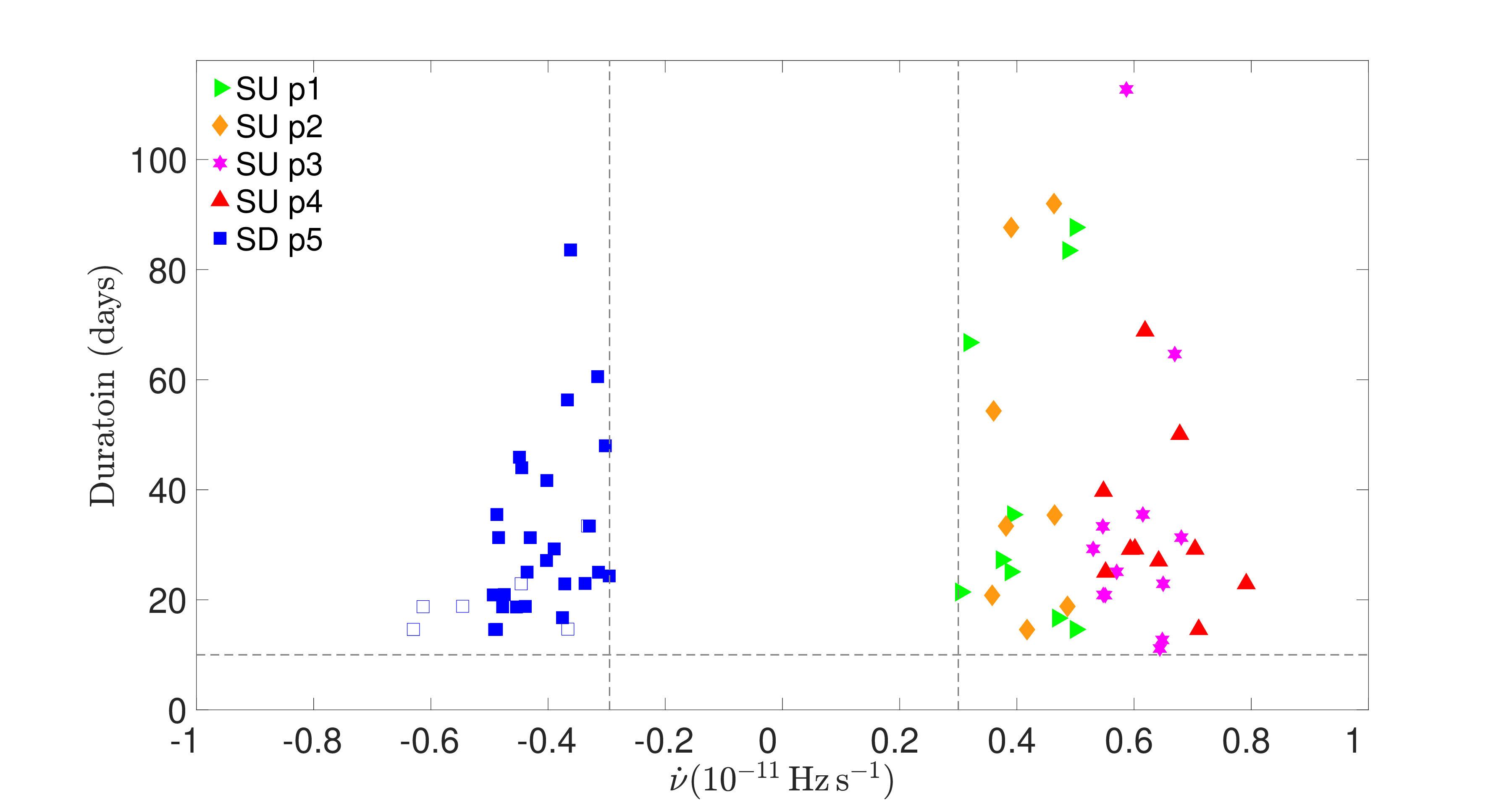}
    \includegraphics[width=1.0\columnwidth]{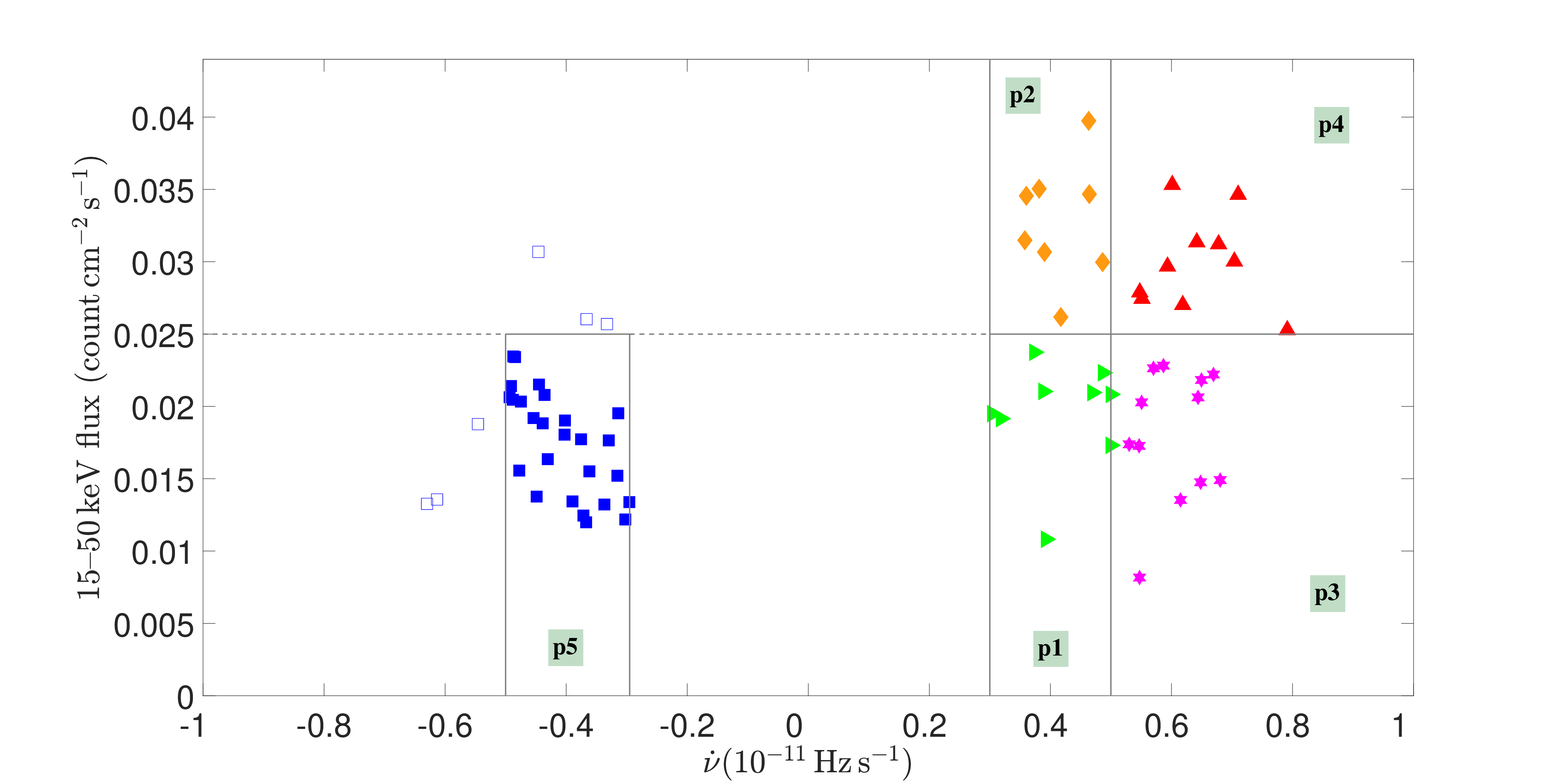}
    \caption{The duration (upper panel) and hard flux (lower panel) versus the spin change rate ($\Dot{\nu}$) for selected SU/SD intervals.
    The dashed lines in the top panel represent the criteria used for selecting SU/SD intervals.
    The regions marked by ``p1''--``p5'' in the bottom panel are defined for subsequent analyses. 
} 
    \label{fig:dn}
\end{figure}

We study the torque-dependent behavior of \cx~based on data monitored by \fer/GBM\footnote{https://gammaray.nsstc.nasa.gov/gbm/science/pulsars/lightcurves/cenx3.html},
\sw/BAT\footnote{https://swift.gsfc.nasa.gov/results/transients/CenX-3/} and \mx/GSC\footnote{http://maxi.riken.jp/star\_data/J1121-606/J1121-606.html}. 
Among these instruments, \fer/GBM monitors the pulse frequency and pulsed flux of dozens of accreting pulsars, 
in the GBM Accreting Pulsar Program \citep{Meegan2009,Mala2020}.
\sw/BAT provides the hard X-ray flux for X-ray transients \citep[15--50\,keV,][]{Krimm2013}, 
while \mx/GSC observes the X-ray sky in the soft X-ray band \citep[2--20\,keV,][]{GSC}. 

The spin evolution history of \cx~is presented in Fig.~\ref{fig:sh}. 
It can be seen that spin-up and spin-down episodes lasting for tens of days alternated in a secular spin-up trend. 
Its spin-up/spin-down rates look smooth, in contrast to typical wind-fed X-ray pulsars, 
such as Vela X-1. To study the spin-up and spin-down episodes separately, 
we identify spin-up/spin-down intervals of continuous trend.
These intervals are required to have an average magnitude of spin change rate higher than $3\times10^{-12}\hzs$ and last longer than 10\,days (roughly 5 orbital cycles).
The selected spin-up (SU) and spin-down (SD) intervals are marked with red and blue in Fig.~\ref{fig:sh}, respectively.

\section{Results}

\begin{figure}
    \centering
    \includegraphics[width=1.0\columnwidth]{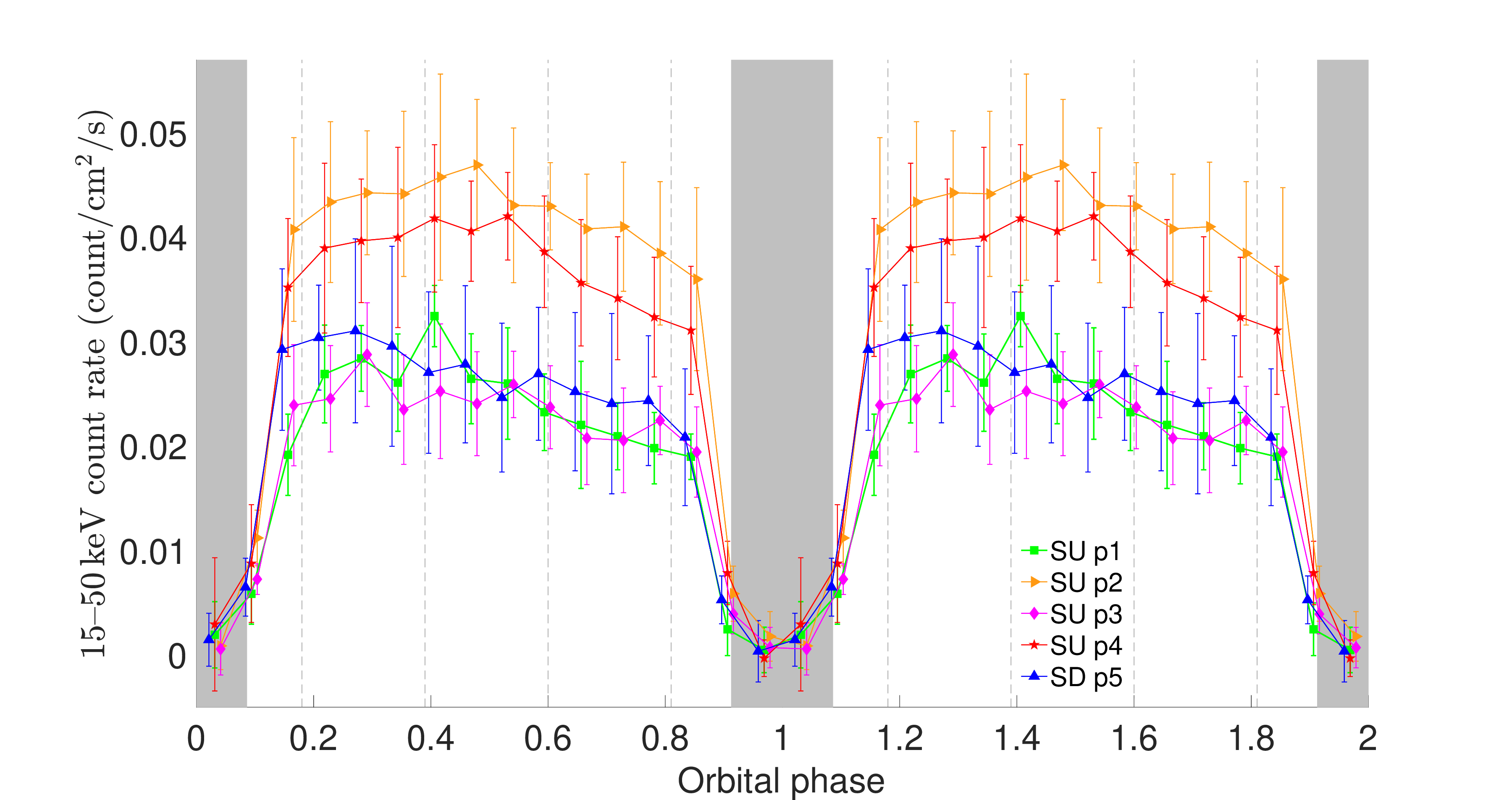} 
    \includegraphics[width=1.0\columnwidth]{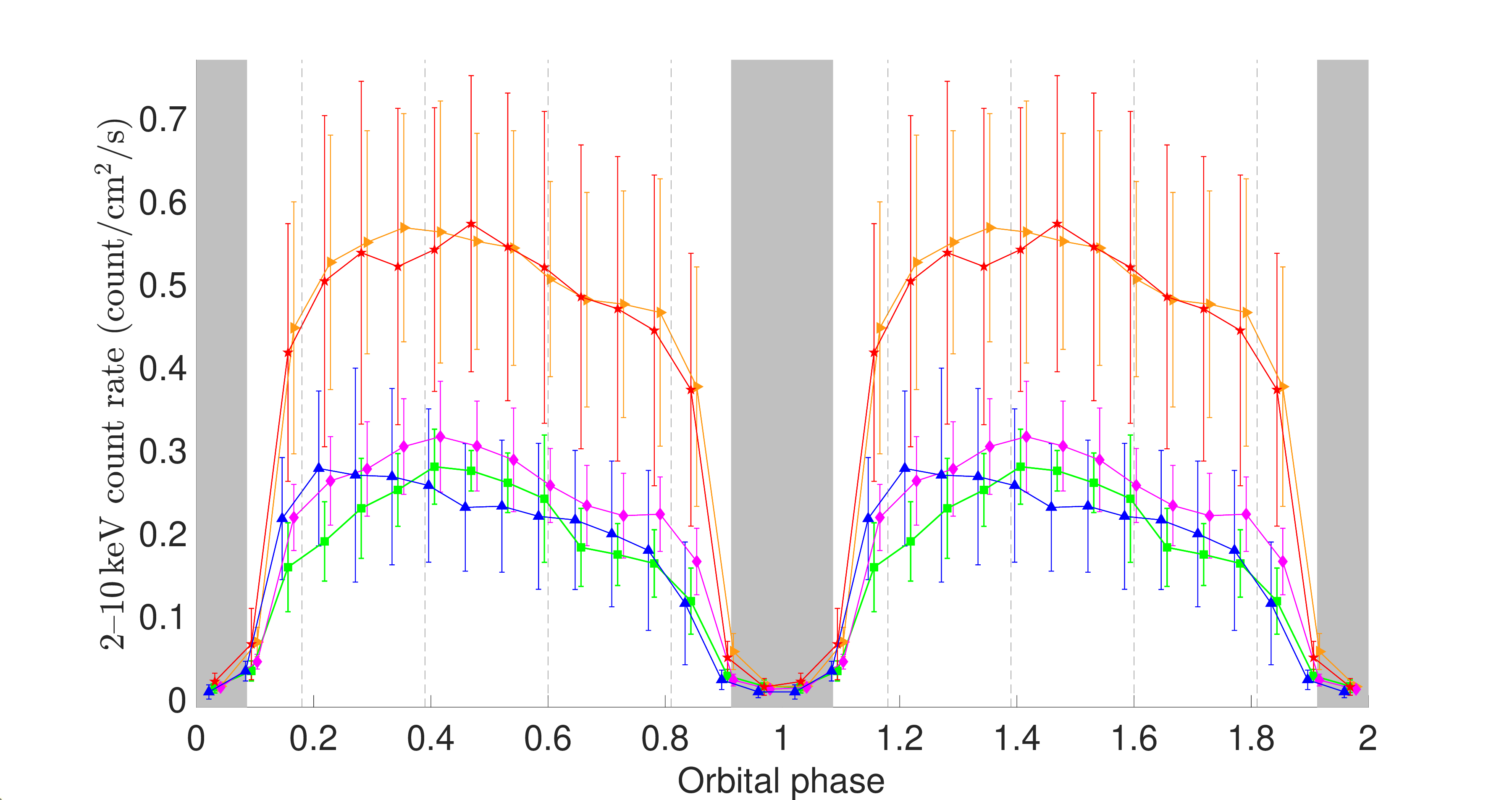} 
    \includegraphics[width=1.0\columnwidth]{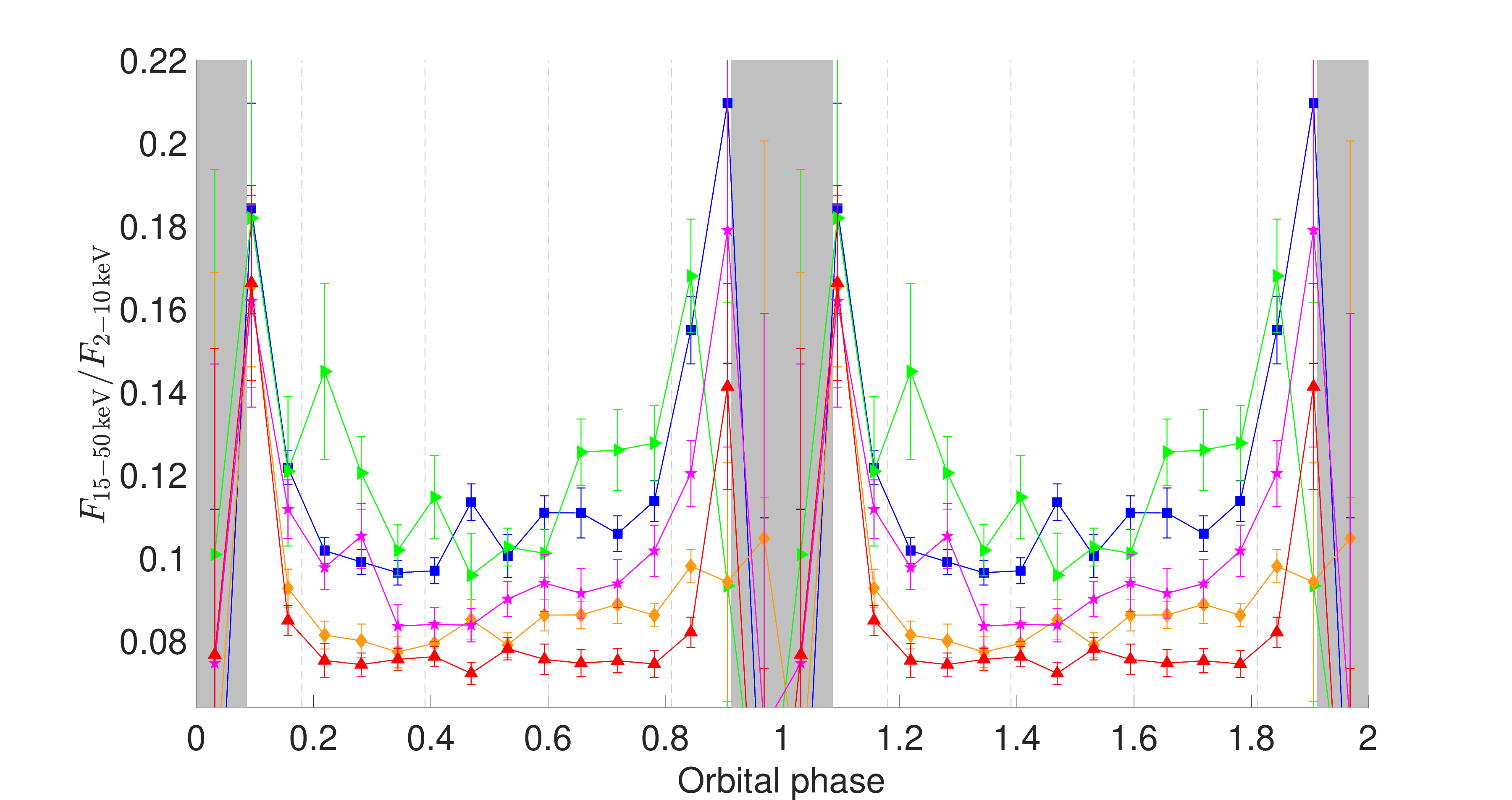} 
    \caption{Orbital profile of \cx~in hard band (15--50\,keV, upper panel) and soft band (2--10\,keV, middle panel), 
    and their hardness ratio for region ``p1''--``p5'' (lower panel). 
    The shaded areas in the plots represent the orbital phases from ingress to egress. 
    }
    \label{fig:opf}
\end{figure}
\begin{figure}
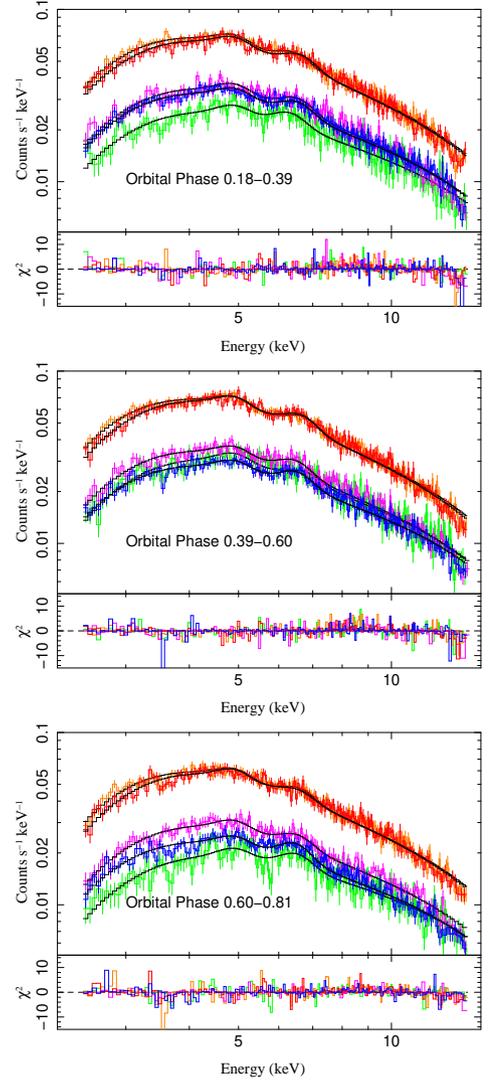

    \centering
    \includegraphics[width=0.55\columnwidth, angle=270]{fig/phi1_cud.eps}
    \includegraphics[width=0.55\columnwidth, angle=270]{fig/phi2_cud.eps}
    \includegraphics[width=0.55\columnwidth, angle=270]{fig/phi3_cud.eps}
    \caption{Extracted spectrum of each orbital phase bin for region ``p1''--``p5'', together with the fitted model (black).
    The upper, middle, and lower panels correspond to orbital phase bins of 0.18--0.39, 0.39--0.60, and 0.60--0.81, respectively. 
    The color schemes are the same as in Fig.~\ref{fig:opf}.}
    \label{fig:spec}
\end{figure}

For each selected interval, we fit a mean spin frequency derivative ($\Dot{\nu}$) and calculate a mean hard X-ray flux (15--50\,keV) from \sw/BAT data. 
The durations of selected intervals are plotted against the spin frequency derivatives ($\Dot{\nu}$) in the upper panel of Fig.~\ref{fig:dn}.
The different symbols in the plot correspond to different SU/SD regions defined in the next paragraph.
We see that most of the durations are within 20--80\,days, 
and the SU intervals have a few more cases longer than 80\,days. 
The mean magnitude of the spin change rate in SD episodes ($|\Dot{\nu}|\sim4.2\times10^{-12}\hzs$) is slightly lower than in SU episodes ($\sim5.3\times10^{-12}\hzs$).

The mean hard X-ray fluxes against the spin frequency derivatives are plotted in the bottom panel of Fig.~\ref{fig:dn}.
No apparent correlation for either SD or SU intervals can be seen, but in general, the SU episodes have more intervals of higher fluxes compared with the SD episodes.
Since the hard X-ray fluxes of most SD intervals are below 0.025\,counts\,cm$^{-2}$\,s$^{-1}$, 
to compare with those SU intervals of similar fluxes and spin change rates, 
we select the SU intervals with fluxes below 0.025\,counts\,cm$^{-2}$\,s$^{-1}$ and $3\times10^{-12}\hzs < \Dot{\nu} < 5\times10^{-12}\hzs$, 
which is marked as ``p1'' in Fig.~\ref{fig:dn}.
The other SU intervals of similar spin change rates but higher fluxes (``p2''), similar fluxes but higher spin change rates (``p3''), 
higher fluxes and higher spin change rates (``p4'') are also marked in Fig.~\ref{fig:dn}.

\subsection{Orbital profiles}

We calculate the orbital profiles of \cx~from \sw/BAT data (15--50\,keV) and \mx/GSC data (2--10\,keV) for the intervals located in the five regions marked in Fig.~\ref{fig:dn}, 
adopting the ephemeris obtained by \citet{Falan2015}. 
The resulting orbital profiles are shown in Fig.~\ref{fig:opf}, together with the hardness ratio profiles. 
For clarity, the orbital profiles of different regions are slightly shifted.
The orbital phase 0 is located at the middle of eclipse, and the orbital phase 0.5 corresponds to the times when the neutron star is observed in front of the companion.
Note that these orbital profiles are averaged over many periods, 
and the orbital profile in one particular period might be quite different.
To reflect the scatter of profiles from individual intervals,
the plotted error bars are the standard deviations of the orbital profile obtained from all intervals, 
weighted by the time duration of every interval.
The error bar of the hardness ratio is calculated by summing over all profiles, utilizing the error propagation formula.
 
The hard band profiles (15--50\,keV) of the high states (``p2'' and ``p4'' in Fig.~\ref{fig:dn}) look similar, 
with a plateau around orbital phases between 0.3 and 0.6.
The hard profiles of the low states (``p1'', ``p3'' and ``p5'' in Fig.~\ref{fig:dn}) also look similar, 
except that the one of the spin-down region ``p5'' shows relatively higher fluxes around orbital phase 0.2.

In the soft band (2--10\,keV), the peaking of the profiles of the high states is more apparent than in the hard band.
On the other hand, the 2--10\,keV profile of the low flux and low spin-up rate region ``p1'' looks quite different from that of the spin-down region ``p5''.
The 2--10\,keV profile of the region ``p1'' is peaked around orbital phase 0.42, 
while that of region ``p5'' rapidly reaches a peak value around phase 0.22 and then gradually declines toward phase 0.8.
The profile of the low flux and high spin-up rate region ``p3'' shows a peak around phase 0.42, similar to that of ``p1''. 
Note that region ``p1'' and ``p5'' have similar mean hard fluxes, but different signs of spin change rates, the magnitudes of which are also similar.

To test the reliability of the difference between the peaks of region ``p1'' and ``p5'', 
we randomly chose half of the corresponding intervals in region ``p1'' and ``p5'' and calculated their peak phases for 1000 times. 
We found that all the resampled peaks of ``p1'' are around phase 0.42, 
while for the resampled peaks of ``p5'', about 99\% are smaller than the orbital phase 0.35 and only 1\% are around phase 0.4. 
Considering that we only resampled half of the available intervals, the real significance of the peaking difference between region ``p1'' and ``p5'' should be higher. 
We also calculated the orbital profiles in 2--4\,keV and 4--10\,keV bands separately and found that they are similar to those in 2--10\,keV band.

The hardness ratio profiles also show some differences between various states.
The two high states of ``p2'' and ``p4'' have the lowest hardness ratios, 
whereas the other states have relatively higher ratios. 
The region ``p1'' seems to have anomalous high ratios around orbital phase 0.25.
Due to very small fluxes during the eclipsing phases, the hardness ratios there show much larger variations.

\subsection{Spectra}

To further explore the spectral differences of \cx~in different torque states, 
we extract the orbital-phase-resolved spectra based on the event data obtained by \mx/GSC \citep{Matsuoka2009}, 
for the five regions separately. 
To avoid absorption by the supergiant atmosphere, the chosen orbital phases are 0.18--0.39, 0.39--0.60, 
and 0.60--0.81, as marked by vertical dashed lines in Fig.~\ref{fig:opf}. 
The corresponding spectra are shown in Fig.~\ref{fig:spec}. 

As can be seen, the spectral differences between the states of high hard fluxes and those of low hard fluxes are more prominent in low energies. 
The spectrum of spin-down region ``p5'' around the orbital phase within 0.18--0.39 has more low-energy photons compared to that of spin-up region ``p1''.
To quantify possible spectral differences, we use a phenomenological model of absorbed power-law plus a Gaussian line ($\mathrm{wabs\times powerlaw + egauss}$) to fit all spectra.
The Gaussian line represents the iron emission line around 6.4\,keV.
The fitting energy band is limited to 2.5--14\,keV, and the width of the Gaussian line is fixed to 0.1\,keV.
The fitting results are listed in Table~\ref{tab:par}, including hydrogen column density ($N_{\rm H}$), 
photon index ($\Gamma$), normalization of power-law, area, center, and equivalent width of the Gaussian model, 
absorption-corrected luminosity, and reduced chi-square ($\chi^2_\nu$) of fitting.
When calculating the luminosity, we assumed a distance of $6.4^{+1.0}_{-1.4}\rm\,kpc$ \citep{Arna2021}.

\begin{figure}
    \centering
    \includegraphics[width=0.85\columnwidth]{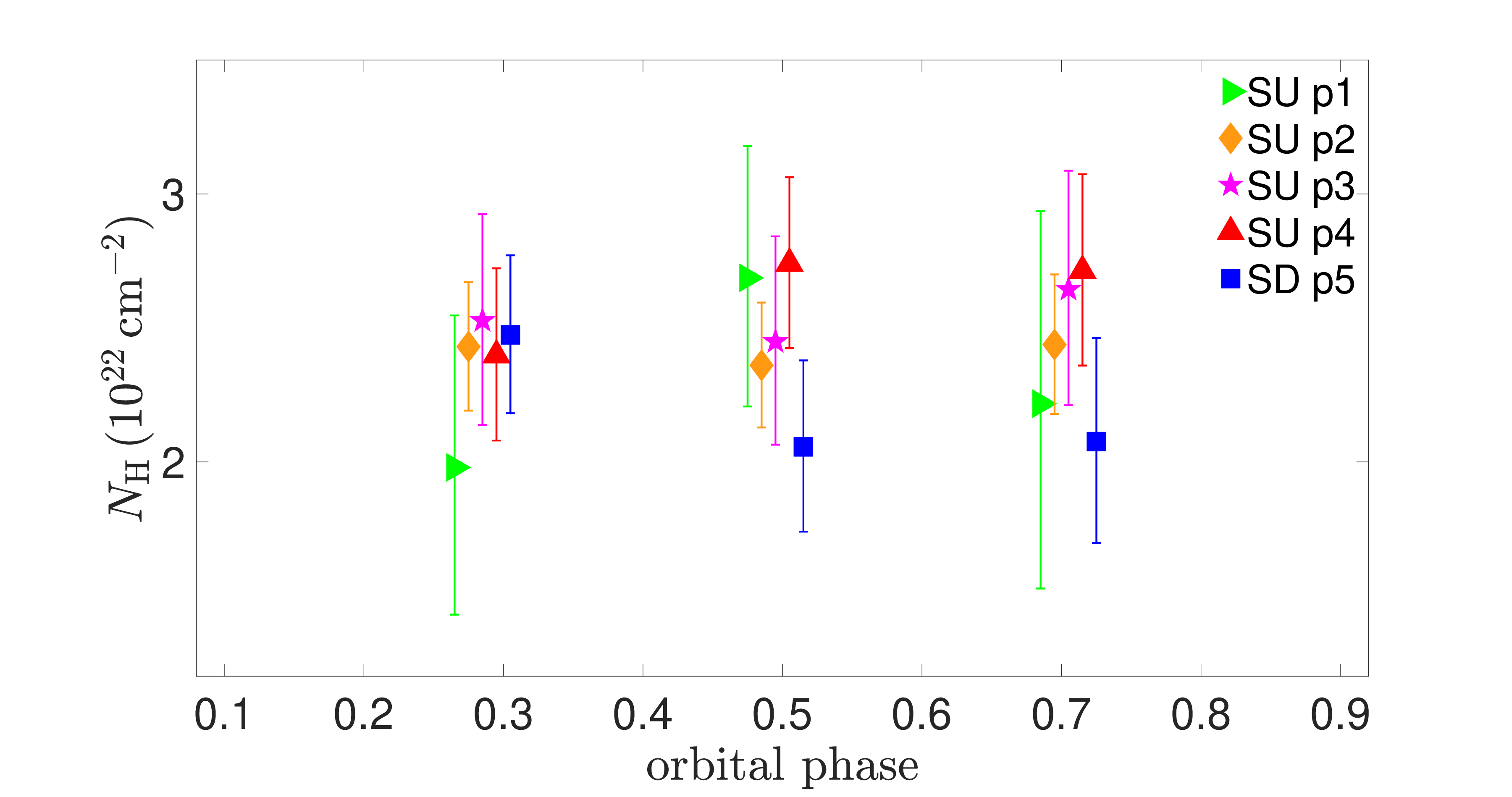}
    \includegraphics[width=0.85\columnwidth]{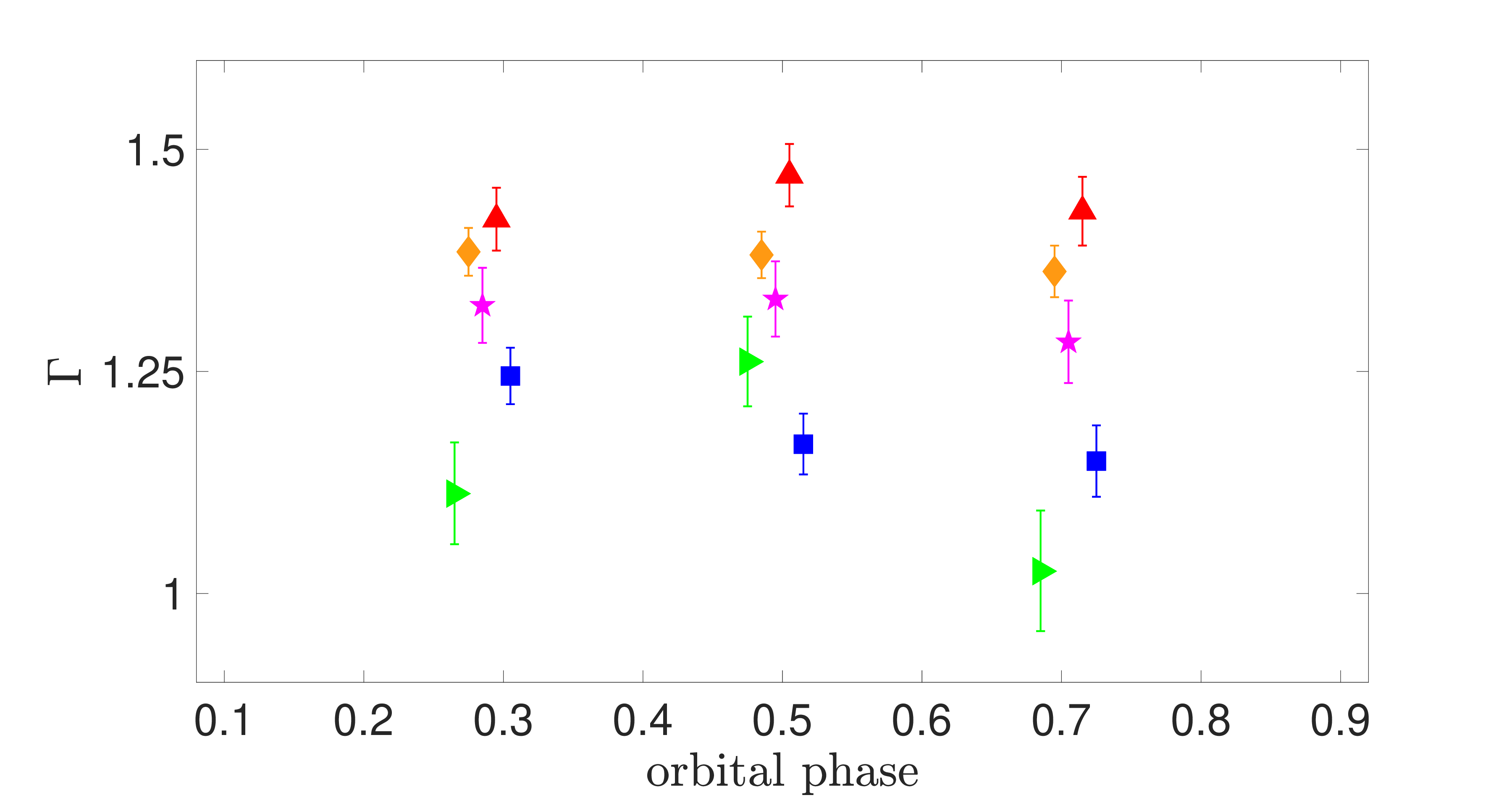}
    \caption{Fitted absorption column densities and photon indices for different torque states.}
    \label{fig:par}
\end{figure}

In Fig.~\ref{fig:par}, we plot the fitting hydrogen column densities and photon indices at around different phases. 
For clarity, the results of different regions are slightly shifted.
It can be seen that the spectral differences are mainly characterized by photon indices:
the higher the fluxes, the larger the spectral indices.
Compared with photon indices, 
the fitted column densities have relatively larger uncertainties, 
and they are similar within error bars.

\begin{table*}
    \renewcommand\arraystretch{1.0}
    \centering
    \caption{Fitting results of an absorbed power-law plus a Gaussian line ($\mathrm{wabs\times powerlaw+egauss}$). Errors are given in 90\% confidence level. The orbital phases of 0.18--0.39, 0.39--0.60, and 0.60--0.80 are represented by phi1, phi2 and phi3, respectively.}
    \begin{threeparttable}
    \begin{tabular}{cccccccccc}
    \hline
    $\phi_{\rm orb}$ & $N_{\rm H}$ & Norm~\tnote{1} & $\Gamma$ & area & center & EW & Luminosity~\tnote{2} & $\chi^2_\nu$ \\
    & ($10^{22}\,\rm cm^{-2}$) & & & ($\rm 10^{-4}\,photon\,s^{-1}\,cm^{-2}$) & (keV) & (eV) & ($10^{37}\egs$) & \\
    \hline
    SU p1 phi1 & $1.98\pm0.56$ & $0.19\pm0.02$ & $1.11\pm0.06$ & $57.17\pm14.80$ & $6.29\pm0.13$ & $245.61\pm44.63$ & $4.96\pm0.57$ & 1.12 \\
    SU p1 phi2 & $2.69\pm0.49$ & $0.29\pm0.03$ & $1.26\pm0.05$ & $39.46\pm13.44$ & $6.67\pm0.13$ & $153.70\pm34.71$ & $6.75\pm0.93$ & 1.15 \\
    SU p1 phi3 & $2.22\pm0.70$ & $0.13\pm0.02$ & $1.03\pm0.07$ & $46.44\pm13.57$ & $6.44\pm0.14$ & $257.71\pm65.10$ & $3.69\pm0.46$ & 1.05 \\
    SU p2 phi1 & $2.43\pm0.24$ & $0.74\pm0.04$ & $1.38\pm0.03$ & $104.56\pm16.45$ & $6.39\pm0.09$ & $193.25\pm17.14$ & $15.35\pm0.92$  & 1.29 \\
    SU p2 phi2 & $2.36\pm0.23$ & $0.73\pm0.04$ & $1.38\pm0.03$ & $103.89\pm14.86$ & $6.62\pm0.09$ & $199.59\pm15.57$ & $15.28\pm1.02$  & 1.15 \\
    SU p2 phi3 & $2.44\pm0.26$ & $0.62\pm0.04$ & $1.36\pm0.03$ & $84.02\pm14.79$ & $6.51\pm0.11$ & $182.01\pm19.12$ & $13.02\pm0.98$  & 1.39 \\
    SU p3 phi1 & $2.53\pm0.39$ & $0.36\pm0.03$ & $1.32\pm0.04$ & $64.61\pm13.54$ & $6.47\pm0.15$ & $223.20\pm30.33$ & $7.78\pm0.64$ & 1.38 \\
    SU p3 phi2 & $2.45\pm0.39$ & $0.36\pm0.03$ & $1.33\pm0.04$ & $65.72\pm13.55$ & $6.40\pm0.11$ & $228.56\pm30.82$ & $7.71\pm0.57$ & 1.22 \\
    SU p3 phi3 & $2.64\pm0.44$ & $0.28\pm0.03$ & $1.28\pm0.05$ & $48.88\pm12.41$ & $6.47\pm0.13$ & $198.88\pm34.80$ & $6.34\pm0.71$ & 0.90 \\
    SU p4 phi1 & $2.40\pm0.32$ & $0.79\pm0.06$ & $1.42\pm0.04$ & $109.71\pm20.38$ & $6.55\pm0.13$ & $209.31\pm28.01$ & $15.98\pm1.37$  & 1.14 \\
    SU p4 phi2 & $2.74\pm0.32$ & $0.89\pm0.07$ & $1.47\pm0.04$ & $116.66\pm20.32$ & $6.56\pm0.10$ & $217.75\pm26.62$ & $17.43\pm1.68$  & 1.10 \\
    SU p4 phi3 & $2.71\pm0.36$ & $0.72\pm0.06$ & $1.43\pm0.04$ & $70.23\pm20.09$ & $6.48\pm0.16$ & $149.27\pm29.48$ & $14.40\pm1.04$  & 1.11 \\
    SD p5 phi1 & $2.47\pm0.29$ & $0.29\pm0.02$ & $1.24\pm0.03$ & $59.54\pm9.42$ & $6.47\pm0.09$ & $218.29\pm30.67$ & $6.75\pm0.55$ & 1.22 \\
    SD p5 phi2 & $2.06\pm0.32$ & $0.22\pm0.02$ & $1.17\pm0.03$ & $57.57\pm9.09$ & $6.46\pm0.08$ & $240.33\pm26.78$ & $5.48\pm0.49$ & 1.01 \\
    SD p5 phi3 & $2.08\pm0.38$ & $0.18\pm0.02$ & $1.15\pm0.04$ & $61.03\pm8.99$ & $6.41\pm0.09$ & $304.32\pm46.15$ & $4.49\pm0.50$ & 1.11 \\
    
    \hline
    \end{tabular}
    \begin{tablenotes}
        \footnotesize
        \item[1] Normalization of the power-law, in units of $\rm photon\,keV^{-1}\,cm^{-2}\,s^{-1}$ at 1\,keV.
        \item[2] Absorption-corrected luminosity within 0.5--20\,keV band.
    \end{tablenotes}
    \end{threeparttable}
    \label{tab:par}
\end{table*}

\section{discussion and conclusion}

We perform orbital timing and spectral analysis of \cx~during different torque states based on the long-term spin history obtained by \fer/GBM, 
and the light curves from \sw/BAT and \mx/GSC.
For intervals with similar hard X-ray fluxes and magnitudes of spin change rate, but with different spin change signs (spin-up or spin-down), 
we find that the 2--10\,keV orbital profile of the spin-up state peaks around phase 0.42 with an approximately symmetric shape, 
in contrast to that of the spin-down state, which reaches a peak value around phase 0.22 and gradually declines toward phase 0.8.
There are also spectral differences between spin-up and spin-down states, characterized by photon indices.

As mentioned in the Introduction, the anti-correlation between torque and luminosity observed in some sources in the spin-down state cannot be explained by the standard disk accretion model. 
The torque-dependent orbital profiles of OAO 1657-415 and Vela X-1 indicated that the torque reversals are somehow related to the accretion flow on the orbital scale, 
not just the scale of the magnetosphere near the neutron star. 
The distinct shapes of the orbital profiles between the spin-up and spin-down states of \cx~with similar fluxes provide another constraint on the torque reversals.
If similar fluxes and similar magnitudes of spin change rate reflect a comparable accretion rate, in order to provide a different orbital profile, 
the accretion flow on the orbital scale should be different for different torque states.

\citet{vanKerkwijk1998} had proposed that the warping instability of the accretion disk due to irradiation may lead to a flipped inner part of the accretion disk,
which is retrograde with respect to the neutron star and can explain the observed spin-down episodes. 
\citet{Raichur2008l} suggested that a precessing disk may explain the flux-dependent orbital profiles of \cx~they found with {\it RXTE} data. 
It is interesting to note that many torque-related properties are consistent with such a model. 
The magnitudes of spin-up rates are generally higher than those of spin-down rates for all three sources we studied (\cx, OAO 1657-415 and \vx),
which was already pointed out by \citet{vanKerkwijk1998}. 
The fluxes of spin-up states are also higher than those of spin-down states for all three sources. 
We also note that the durations of the continuous spin-up/spin-down intervals of \cx~are generally shorter than its possible superorbital periods. 
The reported superorbital periods of Cen X-3 range from about 100\,days to 200\,days \citep{Priedhorsky1983,Tor22}. 
Moreover, the durations themselves indicate a characteristic timescale of the process responsible for the torque reversal, which should be superorbital.
These observed features are consistent with the scenario of a warping disk instability, which may produce a variable prograde/retrograde inner disk lasting for tens of days.

There are also spectral differences between the averaged spectra of spin-up and spin-down states, mainly characteristic by the photon indices.
The fitted photon indices are similar to those obtained from {\it AstroSat} data \citep{Bachhar2022} and those from {\it NuSTAR} data in high state \citep{Dangal2024}.
It is interesting to note that the higher state generally has a larger photon index, 
which is similar to the behavior of the diagonal branch of Be/X-ray pulsars found by \citet{Reig2013}.
Nevertheless, 
considering that absorption column density and photon index are coupled with each other in the spectral fitting, 
and especially the spectra are averaged over many intervals, one cannot be sure whether the fitted similar column densities are real. 
Better spectral data is needed to reveal the real cases.

In summary, 
the distinct orbital difference between spin-up and spin-down states of \cx~with similar fluxes is hard to explain with the standard disk model and indicates that its torque reversals are related to physical processes on the orbital scale.
The durations of continuous spin-up/spin-down trend (tens of days) also point to a superorbital variation. 
The irradiation-driven warping disk instability provides a possible mechanism to produce a flipped inner disk lasting for tens of days. 
In such a scenario, the disk is always there, only the direction of disk is changed. 
That is, the accretion rate should be stable during the torque reversal time. 
On the other hand, if the torque reversal is not due to changing direction of a disk, 
but due to a newly formed disk of opposite rotation, the accretion rate would decrease strongly during the reversal time. 
In principle, these two scenarios could be tested by investigating the luminosity behavior during the reversal time, 
although obtaining the variations of \cx~itself would be a challenge. 

\section*{Acknowledgements}
We thank the referee for his/her thoughtful comments, which considerably improved the paper.
JL acknowledges the support of the National Science Foundation of China (NSFC U1938113).
ZL acknowledges the support of the Startup Foundation of Sanming University (22YG14), 
the Educational Research Project for Young and Middle-aged Teachers of Fujian Provincial Department of Education (JAT220352).
The authors appreciate the valuable discussion with Dr. X. Wang. 
This research has made use of \mx~data provided by {\it RIKEN}, {\it JAXA} and the \mx~team.

\section*{Data Availability}

The data analyzed here are monitored by \fer/GBM, \sw/BAT and \mx/GSC and are all publicly available on their websites.
 



\bibliographystyle{mnras}
\bibliography{cx3bib} 

\begin{thebibliography}{}
\makeatletter
\relax
\def\mn@urlcharsother{\let\do\@makeother \do\$\do\&\do\#\do\^\do\_\do\%\do\~}
\def\mn@doi{\begingroup\mn@urlcharsother \@ifnextchar [ {\mn@doi@}
  {\mn@doi@[]}}
\def\mn@doi@[#1]#2{\def\@tempa{#1}\ifx\@tempa\@empty \href
  {http://dx.doi.org/#2} {doi:#2}\else \href {http://dx.doi.org/#2} {#1}\fi
  \endgroup}
\def\mn@eprint#1#2{\mn@eprint@#1:#2::\@nil}
\def\mn@eprint@arXiv#1{\href {http://arxiv.org/abs/#1} {{\tt arXiv:#1}}}
\def\mn@eprint@dblp#1{\href {http://dblp.uni-trier.de/rec/bibtex/#1.xml}
  {dblp:#1}}
\def\mn@eprint@#1:#2:#3:#4\@nil{\def\@tempa {#1}\def\@tempb {#2}\def\@tempc
  {#3}\ifx \@tempc \@empty \let \@tempc \@tempb \let \@tempb \@tempa \fi \ifx
  \@tempb \@empty \def\@tempb {arXiv}\fi \@ifundefined
  {mn@eprint@\@tempb}{\@tempb:\@tempc}{\expandafter \expandafter \csname
  mn@eprint@\@tempb\endcsname \expandafter{\@tempc}}}

\bibitem[\protect\citeauthoryear{{Arnason}, {Papei}, {Barmby}, {Bahramian}  \&
  {Gorski}}{{Arnason} et~al.}{2021}]{Arna2021}
{Arnason} R.~M.,  {Papei} H.,  {Barmby} P.,  {Bahramian} A.,   {Gorski} M.~D.,
  2021, \mn@doi [\mnras] {10.1093/mnras/stab345}, \href
  {https://ui.adsabs.harvard.edu/abs/2021MNRAS.502.5455A} {502, 5455}

\bibitem[\protect\citeauthoryear{{Ash}, {Reynolds}, {Roche}, {Norton}, {Still}
  \& {Morales-Rueda}}{{Ash} et~al.}{1999}]{Ash1999}
{Ash} T.~D.~C.,  {Reynolds} A.~P.,  {Roche} P.,  {Norton} A.~J.,  {Still}
  M.~D.,   {Morales-Rueda} L.,  1999, \mn@doi [\mnras]
  {10.1046/j.1365-8711.1999.02605.x}, \href
  {https://ui.adsabs.harvard.edu/abs/1999MNRAS.307..357A} {307, 357}

\bibitem[\protect\citeauthoryear{{Bachhar}, {Raman}, {Bhalerao}  \&
  {Bhattacharya}}{{Bachhar} et~al.}{2022}]{Bachhar2022}
{Bachhar} R.,  {Raman} G.,  {Bhalerao} V.,   {Bhattacharya} D.,  2022, \mn@doi
  [\mnras] {10.1093/mnras/stac2901}, \href
  {https://ui.adsabs.harvard.edu/abs/2022MNRAS.517.4138B} {517, 4138}

\bibitem[\protect\citeauthoryear{{Bildsten} et~al.,}{{Bildsten}
  et~al.}{1997}]{Bild1997}
{Bildsten} L.,  et~al., 1997, \mn@doi [\apjs] {10.1086/313060}, \href
  {https://ui.adsabs.harvard.edu/abs/1997ApJS..113..367B} {113, 367}

\bibitem[\protect\citeauthoryear{{Chakrabarty} et~al.,}{{Chakrabarty}
  et~al.}{1997}]{Chak1997}
{Chakrabarty} D.,  et~al., 1997, \mn@doi [\apjl] {10.1086/310666}, \href
  {https://ui.adsabs.harvard.edu/abs/1997ApJ...481L.101C} {481, L101}

\bibitem[\protect\citeauthoryear{{Chodil}, {Mark}, {Rodrigues}, {Seward}  \&
  {Swift}}{{Chodil} et~al.}{1967}]{Chod1967}
{Chodil} G.,  {Mark} H.,  {Rodrigues} R.,  {Seward} F.~D.,   {Swift} C.~D.,
  1967, \mn@doi [\apj] {10.1086/149312}, \href
  {https://ui.adsabs.harvard.edu/abs/1967ApJ...150...57C} {150, 57}

\bibitem[\protect\citeauthoryear{{Dangal}, {Misra}, {Chakradhari}  \&
  {Bhulla}}{{Dangal} et~al.}{2024}]{Dangal2024}
{Dangal} P.,  {Misra} R.,  {Chakradhari} N.~K.,   {Bhulla} Y.,  2024, \mn@doi
  [\mnras] {10.1093/mnras/stad3590}, \href
  {https://ui.adsabs.harvard.edu/abs/2024MNRAS.527.6981D} {527, 6981}

\bibitem[\protect\citeauthoryear{{Falanga}, {Bozzo}, {Lutovinov},
  {Bonnet-Bidaud}, {Fetisova}  \& {Puls}}{{Falanga} et~al.}{2015}]{Falan2015}
{Falanga} M.,  {Bozzo} E.,  {Lutovinov} A.,  {Bonnet-Bidaud} J.~M.,  {Fetisova}
  Y.,   {Puls} J.,  2015, \mn@doi [\aap] {10.1051/0004-6361/201425191}, \href
  {https://ui.adsabs.harvard.edu/abs/2015A&A...577A.130F} {577, A130}

\bibitem[\protect\citeauthoryear{{Finger}, {Wilson}  \& {Fishman}}{{Finger}
  et~al.}{1994}]{Finger1994}
{Finger} M.~H.,  {Wilson} R.~B.,   {Fishman} G.~J.,  1994, in {Fichtel} C.~E.,
  {Gehrels} N.,   {Norris} J.~P.,  eds,  American Institute of Physics
  Conference Series Vol. 304, The Second Compton Symposium. pp 304--308,
  \mn@doi{10.1063/1.45678}

\bibitem[\protect\citeauthoryear{{Giacconi}, {Gursky}, {Kellogg}, {Schreier}
  \& {Tananbaum}}{{Giacconi} et~al.}{1971}]{Giacconi1971}
{Giacconi} R.,  {Gursky} H.,  {Kellogg} E.,  {Schreier} E.,   {Tananbaum} H.,
  1971, \mn@doi [\apjl] {10.1086/180762}, \href
  {https://ui.adsabs.harvard.edu/abs/1971ApJ...167L..67G} {167, L67}

\bibitem[\protect\citeauthoryear{{Klawin}, {Doroshenko}, {Santangelo}, {Ji},
  {Ducci}, {Bu}, {Zhang}  \& {Zhang}}{{Klawin} et~al.}{2023}]{Klawin2023}
{Klawin} M.,  {Doroshenko} V.,  {Santangelo} A.,  {Ji} L.,  {Ducci} L.,  {Bu}
  Q.,  {Zhang} S.-N.,   {Zhang} S.,  2023, \mn@doi [\aap]
  {10.1051/0004-6361/202245181}, \href
  {https://ui.adsabs.harvard.edu/abs/2023A&A...675A.135K} {675, A135}

\bibitem[\protect\citeauthoryear{{Kretschmar} et~al.,}{{Kretschmar}
  et~al.}{2019}]{Kret2019}
{Kretschmar} P.,  et~al., 2019, arXiv e-prints, \href
  {https://ui.adsabs.harvard.edu/abs/2019arXiv190508578K} {p. arXiv:1905.08578}

\bibitem[\protect\citeauthoryear{{Krimm} et~al.,}{{Krimm}
  et~al.}{2013}]{Krimm2013}
{Krimm} H.~A.,  et~al., 2013, \mn@doi [\apjs] {10.1088/0067-0049/209/1/14},
  \href {https://ui.adsabs.harvard.edu/abs/2013ApJS..209...14K} {209, 14}

\bibitem[\protect\citeauthoryear{{Liao}, {Liu}, {Jenke}  \& {Gou}}{{Liao}
  et~al.}{2022a}]{Liao2022o}
{Liao} Z.,  {Liu} J.,  {Jenke} P.~A.,   {Gou} L.,  2022a, \mn@doi [\mnras]
  {10.1093/mnras/stab3561}, \href
  {https://ui.adsabs.harvard.edu/abs/2022MNRAS.510.1765L} {510, 1765}

\bibitem[\protect\citeauthoryear{{Liao}, {Liu}  \& {Gou}}{{Liao}
  et~al.}{2022b}]{Liao2022v}
{Liao} Z.,  {Liu} J.,   {Gou} L.,  2022b, \mn@doi [\mnras]
  {10.1093/mnrasl/slac119}, \href
  {https://ui.adsabs.harvard.edu/abs/2022MNRAS.517L.111L} {517, L111}

\bibitem[\protect\citeauthoryear{{Makishima} et~al.,}{{Makishima}
  et~al.}{1988}]{Maki88}
{Makishima} K.,  et~al., 1988, \mn@doi [\nat] {10.1038/333746a0}, \href
  {https://ui.adsabs.harvard.edu/abs/1988Natur.333..746M} {333, 746}

\bibitem[\protect\citeauthoryear{{Malacaria}, {Jenke}, {Roberts},
  {Wilson-Hodge}, {Cleveland}, {Mailyan}  \& {GBM Accreting Pulsars Program
  Team}}{{Malacaria} et~al.}{2020}]{Mala2020}
{Malacaria} C.,  {Jenke} P.,  {Roberts} O.~J.,  {Wilson-Hodge} C.~A.,
  {Cleveland} W.~H.,  {Mailyan} B.,   {GBM Accreting Pulsars Program Team}
  2020, \mn@doi [\apj] {10.3847/1538-4357/ab855c}, \href
  {https://ui.adsabs.harvard.edu/abs/2020ApJ...896...90M} {896, 90}

\bibitem[\protect\citeauthoryear{{Matsuoka} et~al.,}{{Matsuoka}
  et~al.}{2009}]{Matsuoka2009}
{Matsuoka} M.,  et~al., 2009, \mn@doi [\pasj] {10.1093/pasj/61.5.999}, \href
  {https://ui.adsabs.harvard.edu/abs/2009PASJ...61..999M} {61, 999}

\bibitem[\protect\citeauthoryear{{Meegan} et~al.,}{{Meegan}
  et~al.}{2009}]{Meegan2009}
{Meegan} C.,  et~al., 2009, \mn@doi [\apj] {10.1088/0004-637X/702/1/791}, \href
  {https://ui.adsabs.harvard.edu/abs/2009ApJ...702..791M} {702, 791}

\bibitem[\protect\citeauthoryear{{Mihara} et~al.,}{{Mihara} et~al.}{2011}]{GSC}
{Mihara} T.,  et~al., 2011, \mn@doi [\pasj] {10.1093/pasj/63.sp3.S623}, \href
  {https://ui.adsabs.harvard.edu/abs/2011PASJ...63S.623M} {63, S623}

\bibitem[\protect\citeauthoryear{{Nelson} et~al.,}{{Nelson}
  et~al.}{1997}]{Nelson1997}
{Nelson} R.~W.,  et~al., 1997, \mn@doi [\apjl] {10.1086/310936}, \href
  {https://ui.adsabs.harvard.edu/abs/1997ApJ...488L.117N} {488, L117}

\bibitem[\protect\citeauthoryear{{Priedhorsky} \& {Terrell}}{{Priedhorsky} \&
  {Terrell}}{1983}]{Priedhorsky1983}
{Priedhorsky} W.~C.,  {Terrell} J.,  1983, \mn@doi [\apj] {10.1086/161406},
  \href {https://ui.adsabs.harvard.edu/abs/1983ApJ...273..709P} {273, 709}

\bibitem[\protect\citeauthoryear{{Raichur} \& {Paul}}{{Raichur} \&
  {Paul}}{2008}]{Raichur2008l}
{Raichur} H.,  {Paul} B.,  2008, \mn@doi [\mnras]
  {10.1111/j.1365-2966.2008.13251.x}, \href
  {https://ui.adsabs.harvard.edu/abs/2008MNRAS.387..439R} {387, 439}

\bibitem[\protect\citeauthoryear{{Reig} \& {Nespoli}}{{Reig} \&
  {Nespoli}}{2013}]{Reig2013}
{Reig} P.,  {Nespoli} E.,  2013, \mn@doi [\aap] {10.1051/0004-6361/201219806},
  \href {https://ui.adsabs.harvard.edu/abs/2013A&A...551A...1R} {551, A1}

\bibitem[\protect\citeauthoryear{{Schreier}, {Levinson}, {Gursky}, {Kellogg},
  {Tananbaum}  \& {Giacconi}}{{Schreier} et~al.}{1972}]{Schreier1972}
{Schreier} E.,  {Levinson} R.,  {Gursky} H.,  {Kellogg} E.,  {Tananbaum} H.,
  {Giacconi} R.,  1972, \mn@doi [\apjl] {10.1086/180896}, \href
  {https://ui.adsabs.harvard.edu/abs/1972ApJ...172L..79S} {172, L79}

\bibitem[\protect\citeauthoryear{{Tjemkes}, {Zuiderwijk}  \& {van
  Paradijs}}{{Tjemkes} et~al.}{1986}]{Tjemkes1986}
{Tjemkes} S.~A.,  {Zuiderwijk} E.~J.,   {van Paradijs} J.,  1986, \aap, \href
  {https://ui.adsabs.harvard.edu/abs/1986A&A...154...77T} {154, 77}

\bibitem[\protect\citeauthoryear{{Torregrosa}, {Rodes-Roca}, {Torrej{\'o}n},
  {Sanjurjo-Ferr{\'\i}n}  \& {Bernab{\'e}u}}{{Torregrosa} et~al.}{2022}]{Tor22}
{Torregrosa} {\'A}.,  {Rodes-Roca} J.~J.,  {Torrej{\'o}n} J.~M.,
  {Sanjurjo-Ferr{\'\i}n} G.,   {Bernab{\'e}u} G.,  2022, \mn@doi [\rmxaa]
  {10.22201/ia.01851101p.2022.58.02.15}, \href
  {https://ui.adsabs.harvard.edu/abs/2022RMxAA..58..355T} {58, 355}

\bibitem[\protect\citeauthoryear{{Tsunemi}, {Kitamoto}  \& {Tamura}}{{Tsunemi}
  et~al.}{1996}]{Tsunemi1996}
{Tsunemi} H.,  {Kitamoto} S.,   {Tamura} K.,  1996, \mn@doi [\apj]
  {10.1086/176652}, \href
  {https://ui.adsabs.harvard.edu/abs/1996ApJ...456..316T} {456, 316}

\bibitem[\protect\citeauthoryear{{van Kerkwijk}, {Chakrabarty}, {Pringle}  \&
  {Wijers}}{{van Kerkwijk} et~al.}{1998}]{vanKerkwijk1998}
{van Kerkwijk} M.~H.,  {Chakrabarty} D.,  {Pringle} J.~E.,   {Wijers}
  R.~A.~M.~J.,  1998, \mn@doi [\apjl] {10.1086/311346}, \href
  {https://ui.adsabs.harvard.edu/abs/1998ApJ...499L..27V} {499, L27}

\bibitem[\protect\citeauthoryear{{van der Meer}, {Kaper}, {van Kerkwijk},
  {Heemskerk}  \& {van den Heuvel}}{{van der Meer} et~al.}{2007}]{vdm2007}
{van der Meer} A.,  {Kaper} L.,  {van Kerkwijk} M.~H.,  {Heemskerk} M.~H.~M.,
  {van den Heuvel} E.~P.~J.,  2007, \mn@doi [\aap]
  {10.1051/0004-6361:20066025}, \href
  {https://ui.adsabs.harvard.edu/abs/2007A&A...473..523V} {473, 523}

\makeatother
\end{thebibliography}








\bsp	
\label{lastpage}
\end{document}